# Active coupling control in densely packed subwavelength waveguides via dark mode

Haim Suchowski[1,*], Michael Mrejen[1,*], Taiki Hatakeyama[1,*], Chihhui Wu[1] Liang Feng[1], Kevin O'Brien[1], Yuan Wang[1], Xiang Zhang[1,2,†]

[1]NSF Nanoscale Science and Engineering Center (NSEC), University of California, Berkeley

3112 Etcheverry Hall, UC Berkeley, CA 94720, USA

[2]Materials Sciences Division, Lawrence Berkeley National Laboratory, 1 Cyclotron Road, Berkeley, CA 94720, USA

* Equal contribution

[†]To whom correspondence should be addressed. E-mail: xiang@berkeley.edu

# First submitted on August 13[th] 2014

The ever growing need for energy-efficient and fast communications is driving the development of highly integrated photonic circuits where controlling light at the nanoscale becomes the most critical aspect of information transfer [1]. Directional couplers [2], two interacting optical waveguides placed in close proximity, are important building blocks in these integrated photonics circuits and have been employed as optical modulators and switches for high speed communication [3-5], data processing and integrated quantum operations [6]. However, active control over the coupling between closely packed waveguides is highly desirable and yet remains a critical barrier towards ultra small footprint devices. A general approach to achieve active control in waveguide systems is to exploit optical nonlinearities enabled by a strong control pulse [7-10]. However these devices suffer from the nonlinear absorption induced by the intense control pulse as the signal and its control propagate in the same waveguide [7,10]. Here we experimentally demonstrate a unique scheme based on adiabatic elimination (AE) concept that effectively manipulates the coupling between densely packed waveguides. We demonstrate active coupling control between two closely packed waveguides by tuning the mode index of an in-between decoupled waveguide. This is achieved via a dark mode and thus leaves the signal unaffected by the induced losses. Such a scheme is a promising candidate for ultra-dense integrated nano-photonics such as on-chip ultrafast modulators and tunable filters for optical communication and quantum computing.

Two coupled modes are a cornerstone in many research fields – from the dynamics of spin-half systems in nuclear magnetic resonances through two-level systems in atomic and molecular physics to polarization optics and directional couplers [2,11,12]. The dynamics in these systems are generally dictated by two independent quantities: the coupling strength and the phase difference between the interacting states or modes. While active control of these quantities is available in the atomic systems, it remains a challenge in photonics realizations. Quantum control theory (QCT) offers the mathematical framework to decompose multi-levels systems represented by multidimensional Hilbert spaces into *controllable* two coupled mode systems [13]. For instance, in three level atomic systems, several schemes were proposed to reduce the full 8-dimensional Hilbert space into the 3-dimensional space of atomic two level systems [13,14]. One example is the atomic stimulated rapid adiabatic passage (STIRAP) scheme [15], which was recently shown in coupled waveguide systems [16-18]. However this scheme cannot provide a dynamic control of energy transfer between the two outer waveguides, and moreover requires long propagation distance along the waveguides which makes device scaling down difficult.

Here we propose and experimentally demonstrate a unique scheme to actively control the coupling among waveguides for densely integrated photonics based on adiabatic elimination (AE). Analogous to atomic systems [15,19], AE is achieved by applying a decomposition on a three waveguide coupler, where the two outer waveguides serve as an effective two-mode system analogous to the ground and excited states in an atomic three level system, and the middle waveguide is the equivalent to the intermediate 'dark' state. We experimentally show the AE decomposition scheme in nanowaveguides and its capability of active control of the coupling between two waveguides by manipulating the mode index of a decoupled middle one. As the AE procedure separates the signal (information) from the control, it reduces significantly two-photon absorption (TPA) and TPA-induced free carrier absorption experienced by the signal, which are major obstacles in Silicon photonics [10]. In addition, our analysis also includes the inherent higher order coupling, which is significant when packing is subwavelength, allowing control capabilities beyond what can be achieved in atomic physics. This new class of AE-based nanophotonic devices lays down the foundation for ultra-dense integrated photonic circuits.

The evolution of the electrical fields in three coupled mode systems can be described by the general equations [28,29]: $\partial_z A_i = i \left( \Delta\beta_{ij} A_i + \sum_{j \neq i} V_{ij} A_j \right)$, $i, j = 1, 2, 3$, where $A_i$ are the electromagnetic fields of the different modes, $V_{ij}$ is the coupling strength between waveguides $i$ and $j$, and $\Delta\beta_{ij}$ is the mode propagation constant difference, $\Delta\beta_{ij} = \beta_i - \beta_j = \frac{\omega}{c}(n_i^{eff} - n_j^{eff})$, where $\omega$ is the frequency of the light, $c$ is the speed of light and $n_i^{eff}$ is the effective mode index of waveguide $i$. The adiabatic elimination (AE) procedure in waveguides, in analogy to atomic physics, relies on a strong coupling between nearby waveguides that exhibit a very large mode index mismatch between them ($|\Delta\beta_{12}|, |\Delta\beta_{23}| \gg V_{12}, V_{23}$). Since each of the two consecutive coupling processes is greatly mismatched, the amplitude of the middle waveguide, $A_2$ will oscillate very rapidly in comparison to the slow varying amplitudes $A_1$ and $A_3$. As a result, the amplitude of the intermediate waveguide cannot build up significantly and thus remains as in its initial value (i.e. $|A_2(z)| \approx |A_2(z = 0)|$. Fig. 1 shows a schematic comparison between the evolution of identical three-mode system and the AE evolution both in the atomic physics realization and in directional waveguides.

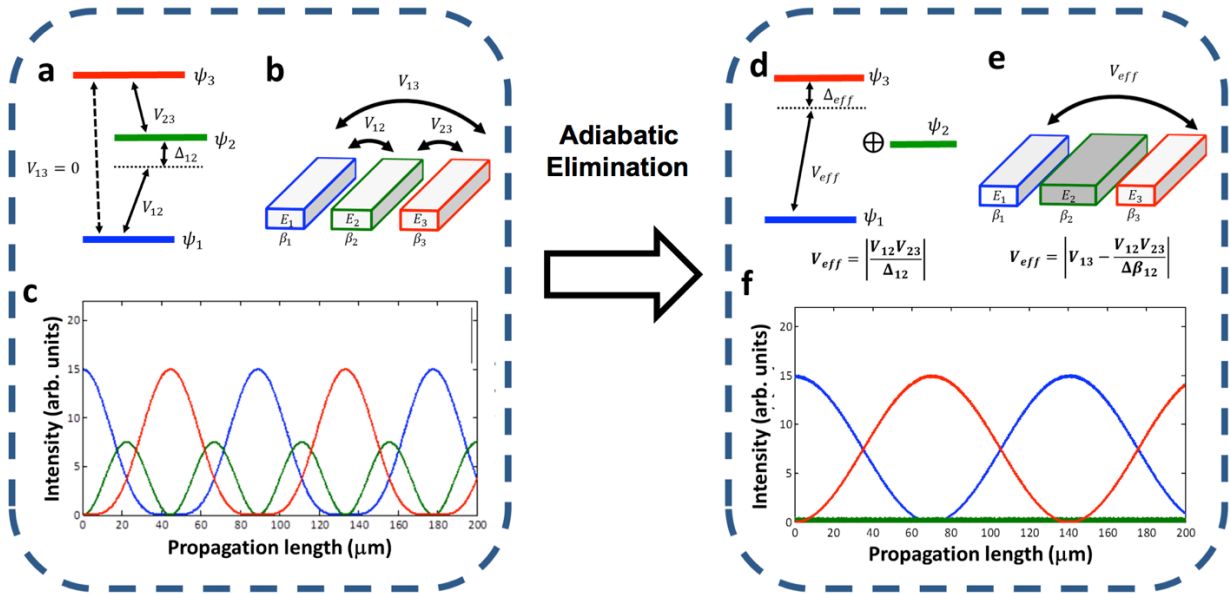

**Figure 1 – Adiabatic elimination (AE) scheme in atomic physics and optical waveguides.** A general three-coupled mode system can be realized in (a) atomic systems and in (b) three waveguide coupler. These systems share equivalent dynamics where the time evolution of the population of electrons in each level is analogous to the electric field propagation in each waveguide. In both cases the evolution is dictated by the couplings between the modes, $V_{ij}$ and by the detunings $\Delta_{ij}$ or the difference between propagation constants $\Delta\beta_{ij}$. (c) The intensity evolution in three identical waveguides, where $V_{12} = V_{23}$ and the light is injected in waveguide 1. This evolution is equivalent to on-resonant three level atomic interactions, where all the electrons are initially in the ground state. As seen, all the waveguides have a significant light intensity throughout the propagation as a result of the couplings. (d) AE process in the atomic system relies on a strong coupling between nearby levels that exhibit a very large detuning between them ($\Delta_{12}, \Delta_{23} \gg V_{12}, V_{23}$).

**Since each of the two coupling processes is greatly detuned, the amplitude of the intermediate level oscillates very rapidly in comparison to the slow varying population in the other levels resulting in no significant build up and remaining at its initial value. The three-level system is thus reduced to an effective 2-level system with an effective coupling $V_{eff} = \left|\frac{V_{12}V_{23}}{\Delta\beta_{12}}\right|$ between the ground and the excited states, with the intermediate level being a 'dark' state. (e) AE analogue in optical waveguides. The outer waveguides become an effective 2-mode coupler with $V_{eff} = V_{13} - \frac{V_{12}V_{23}}{\Delta\beta_{12}}$ and the middle waveguide is a 'dark mode'. Importantly, the coupling between the two outer waveguides is now controllable through $\Delta\beta_{12}$. (f) When AE conditions are met, the light injected in waveguide 1 propagates only in the outer waveguides 1 and 3. The middle waveguide is effectively eliminated, as its energy buildup remains very low during the entire propagation.**

Once AE conditions have been imposed, the effective coupling coefficient in this reduced two level dynamics between the two outer waveguides becomes $V_{eff} = \sqrt{\left(V_{13}^* + \frac{V_{23}^* V_{12}^*}{\Delta\beta_{12}}\right)\left(V_{13} - \frac{V_{12}V_{23}}{\Delta\beta_{23}}\right)}$, and energy can be efficiently transferred between the outer waveguides via the middle one when the outer waveguides have the same mode index ($|\Delta\beta_{12} + \Delta\beta_{23}|z \ll 1$). This expression reveals that the coupling between the outer waveguides depends not only on the couplings in the original three-waveguide system, but also on the difference in the mode propagation constants between the outer waveguides and the middle waveguide even though the latter is 'eliminated'. In other words, one can control the coupling between the outer waveguides if the AE conditions can be met, for example, by changing the effective refractive index of the middle waveguide. Moreover, the efficiency of conversion from $A_1$ to $A_3$ does not change even when the middle waveguide is extremely lossy as we show in the Supplementary Materials (S1). An important parameter that characterizes the degree of elimination in such an adiabatic process is $\frac{V_{12}}{\Delta\beta_{12}}$, which should be much lower than unity in order to ensure AE evolution. Also, it is worth noting that the propagation constant mismatch of the effective two-mode system is modified as well: $\Delta\beta_{eff} = \Delta\beta_{12} + \Delta\beta_{23} + \frac{|V_{12}|^2}{2\Delta\beta_{12}} + \frac{|V_{23}|^2}{2\Delta\beta_{23}} = \Delta\beta_{TP} + \Delta\beta_S$, where in addition to the conventional phase-mismatch terms, two more terms, equivalent to the Stark-shift terms in atomic physics, influence the evolution. This atomic physics analogue for active control with a nanoscale footprint expands earlier theoretical analyses of three waveguide couplers [20-23], and brings new insight for sub-wavelength Silicon devices. It should also be noted that in contrast to the waveguide analogue to STIRAP, where the couplings between the waveguides vary adiabatically along the propagation, in the AE scheme the couplings between the waveguides remain fixed, and the adiabaticity resides in the slow oscillating energy transfer between waveguide 1 and 3 as compared with the very fast oscillations between 1 and 2 and 2 and 3.

We experimentally demonstrate this unique AE in directional waveguides using Silicon on Insulator (SOI) platform. The sample consists of three waveguides, where two identical outer waveguides have the same width, thus the same effective refractive index and a varying middle waveguide width.  The middle waveguide was designed wider than the outer waveguides in order to meet the AE conditions, while using same width for the control sample. In all the configurations the gap between the outer waveguides was kept constant.

We observed that in the AE configuration, the dynamics is indeed decomposed into an effective two-mode system and a dark mode. When light was coupled to an outer waveguide (#1), it remains only in the two outer waveguides along the entire propagation (Fig. 2(c)). On the other hand, when light is injected to the middle waveguide, it remains trapped along the propagation, without coupling to the outer waveguides (Fig. 2(d)). In all cases, properly designed grating couplers and spot size converters have been employed. This is in clear contrast to the controls where all the waveguides participate in the evolution for both injections Fig. 2(a)-(b). Also, we have confirmed these dynamics with both far-field and near field measurements. Those are in a very good agreement with the numerical simulation presented in Fig. S4-S5.

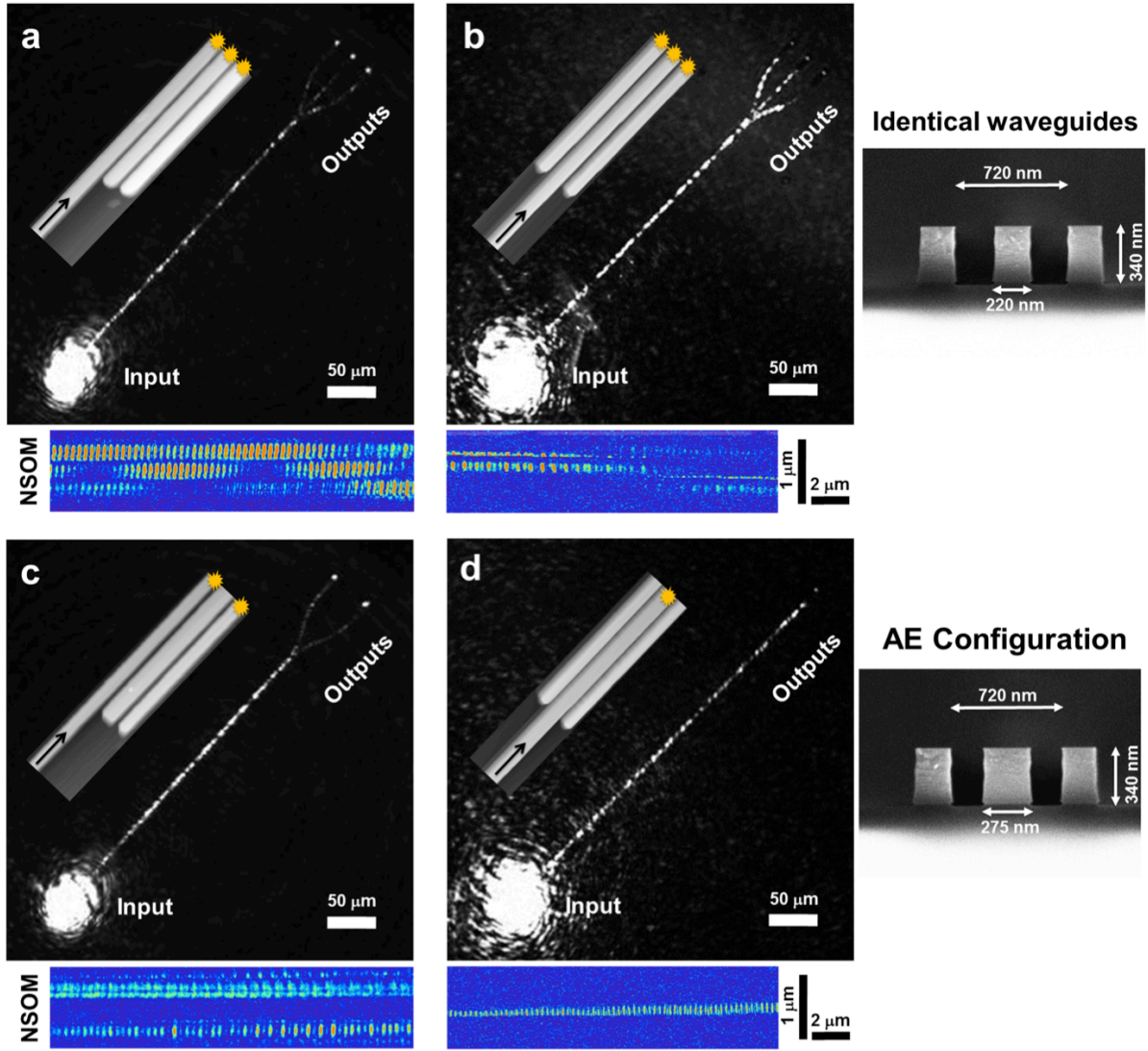

**Figure 2 – Experimental observation of adiabatic elimination compared to ordinary three identical waveguides. (a)-(b) Far-field and NSOM measurements in a three identical waveguides configuration. Due to the coupling between the waveguides, light appears in all waveguides along the propagation, regardless if the input of the light is injected to (a) the outer or (b) the middle waveguide. (c)-(d) Experimental observation of adiabatic elimination (AE) in an AE configuration with a wider middle waveguide. We observe the '2+1' decomposition of three coupled modes into (c) an effective two coupled mode system of the outer waveguides, and (d) a dark middle waveguide. (c) Light is coupled to an outer waveguide in an AE configuration. Far field image shows that only the outer waveguides emit light at the output. The near-field image confirms absence of light in the middle waveguide along the entire propagation. A full NSOM scan showing a complete inversion can be viewed in S4. (d) Light is coupled to the middle waveguide in an AE configuration. Only the middle waveguide emits light at the output without coupling into nearby outer waveguides. The near-field image shows that no light leaks out from the middle waveguide. Inset AFM images shows each of the three waveguides with the inputs and the outputs. The SEM images show the cross section of the fabricated waveguides for the AE configuration and for the identical waveguides configuration. We used a CW laser with $\lambda = 1310\ nm$, the width of the outer waveguides was fixed to $W_1 = W_3 = 220\ nm$, where for the AE configuration $W_2 = 275\ nm$, and for the control waveguide**

**configuration $W_2 = 220\ nm$. The height of the waveguides is 340 nm and the gap between the outer waveguides kept at 720 nm for both cases. The propagation length is 300 $\mu m$ for all configurations.**

The AE process in three coupled waveguides system can be understood using a band diagram of the system's eigenmodes as a function of the middle waveguide width $W_2$ (Fig.3). We observe that in the control configuration the eigenmodes of the coupled system involve necessarily all of the waveguides. However the AE regime, which occurs when the mode index of the middle waveguide differs significantly from the mode index of the outer waveguides, gives rise to an increasingly decoupled eigenmode involving the middle waveguide only. The two other eigenmodes involve only the outer waveguides in a symmetric and anti-symmetric fashion, similar to the conventional two-mode coherent coupler. Hence, in the AE regime, the dynamics can be split into 2+1 (effective two-level and a 'dark state', respectively) dynamical spaces. This decomposition occurs for both TM (shown in Fig. 3) and TE polarizations (shown in S2). Moreover the AE regime is very robust and broadband (see S3). Also, we note that in the case where the middle waveguide is narrower than the outer waveguide, i.e. its mode index is smaller than the outer modes indices, another regime of AE dynamics exists with possible higher effective couplings (as the two contributions in the effective coupling expression will add up constructively), however, with a risk of wavelength cut-off.

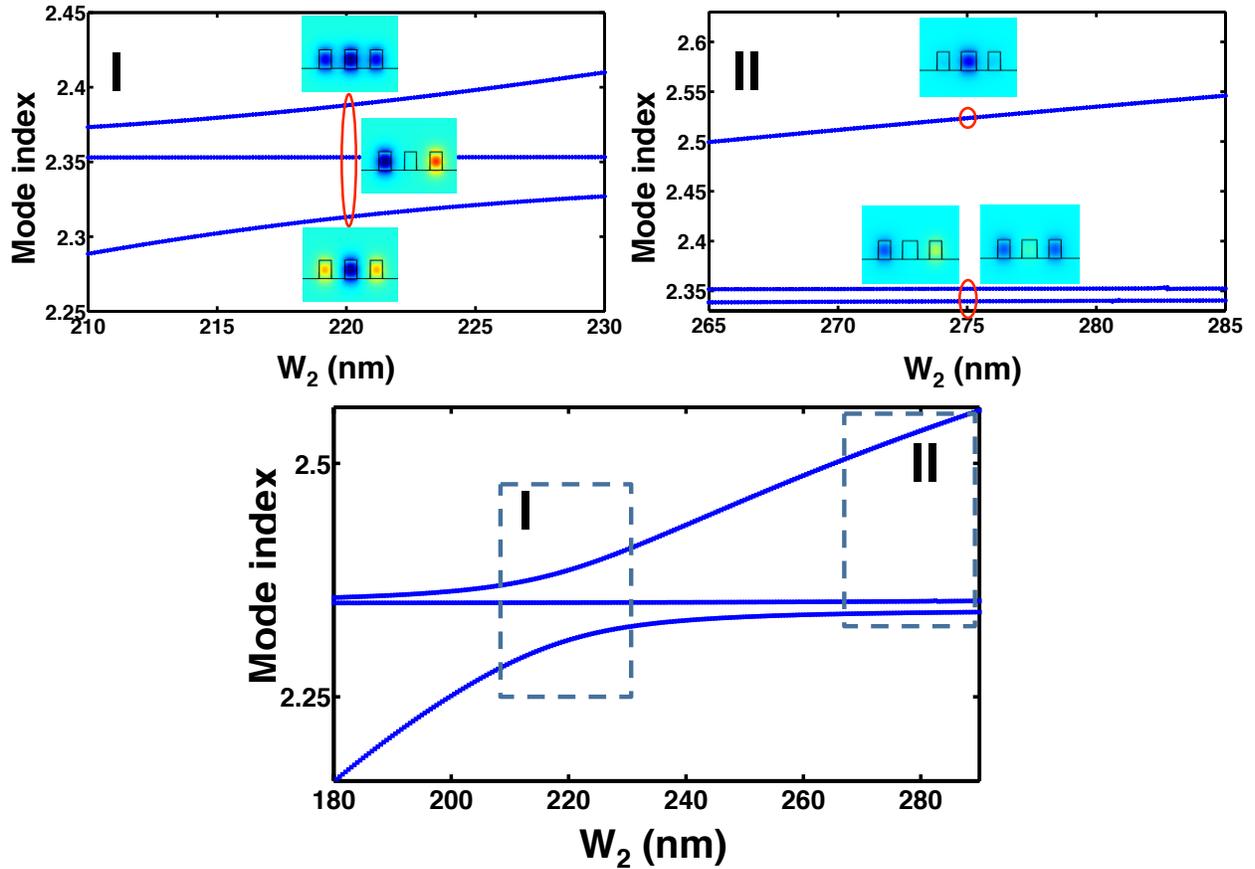

**Figure 3 – Band diagram of a three-mode coupler with varying middle waveguide width, $W_2$. The band diagram shows the electric field eigenvalues and eigenmodes of the system. (Inset I) All the three waveguides are identical, $W_1 = W_2 = W_3 = 220$ nm and are mixed in the eigenmodes of the system. In this configuration, an input in one of the waveguides is projected necessarily on all the waveguides. (Inset II) AE regime. The AE is reached when $W_2$ is larger and our sample (red circles, $W_2 = 275$ nm) is in this regime. Here, an eigenmode involving only the middle waveguide emerges and becomes less coupled, since the coupling is proportional to $1/\Delta n$ where $\Delta n = n_1^{eff} - n_2^{eff}$, leading to the elimination of the middle waveguide. The two remaining eigenmodes are symmetric and anti-symmetric superpositions of the outer waveguides as in a conventional two-mode coherent coupler. An input in either of the outer waveguides will be projected only onto those two eigenmodes, resulting in an effective two-waveguide system evolution, while the middle waveguide remains 'dark' along the propagation. On the other hand, an input injected in the middle waveguide is projected only onto the dark mode of the system and will stay 'trapped' in the middle waveguide without transferring energy to the adjacent ones.**

We now apply the AE scheme to dynamically control the effective coupling between the outer waveguides by changing the mode index of the decoupled middle waveguide. Taking advantage of the large nonlinear Kerr coefficient of Silicon [24] combined with the high light confinement enabled by the sub-wavelength cross section of Silicon photonic waveguides [25], we obtained a localized index change in the middle waveguide.

Using a single-shot nonlinear technique [26], we couple the signal to one of the outer waveguides while the pump beam is coupled to the middle waveguide (Fig. 4a). The AE evolution confines the pump beam to the middle waveguide along the propagation thus ensuring no leakage to the outer waveguides. We insure proper synchronization between the signal and the pump by monitoring the Sum Frequency Generation (SFG) in a BBO crystal (Fig. 4b and S5). We clearly observe that the signal collected is modulated in the presence of the pump (Fig. 4c). The modulation originates from the nonlinear change of refractive index in the middle waveguide as the pump beam propagates. This change, which increases the phase-mismatch between the middle and the outer waveguides ($n_{middle} = n^0_{middle} + n_2 I_{pump} \rightarrow \Delta\beta_{12} = \Delta\beta^0_{12} + n_2 I_{pump}/\lambda$ ), in turn affects the effective coupling between waveguides #1 and #3 which, finally, leads to a change in the inversion length ($L_{inv}$), manifested as a change of output intensity in a waveguide of a given length ($\Delta\beta_{12}$ increases$\rightarrow \Delta V_{eff}$ decreases $\rightarrow L_{inv}$ increases).

Our numerical simulations of the coupled nonlinear Schrodinger equations show that this result discussed above is consistent with a change of index of $\Delta n = 7.15 \pm 0.25 \cdot 10^{-3}$ in the middle waveguide with a pulse energy of 66 ± 6 pJ. We also confirm that in the presence of the pump in the middle waveguide the inversion length in the outer waveguides increases. We have checked the effect as a function of pump-signal delay and confirmed that the effect observed in the synchronized measurements stems from an ultrafast change of index in the middle waveguide and that TPA generated free carriers do no play a significant role in the process [10].

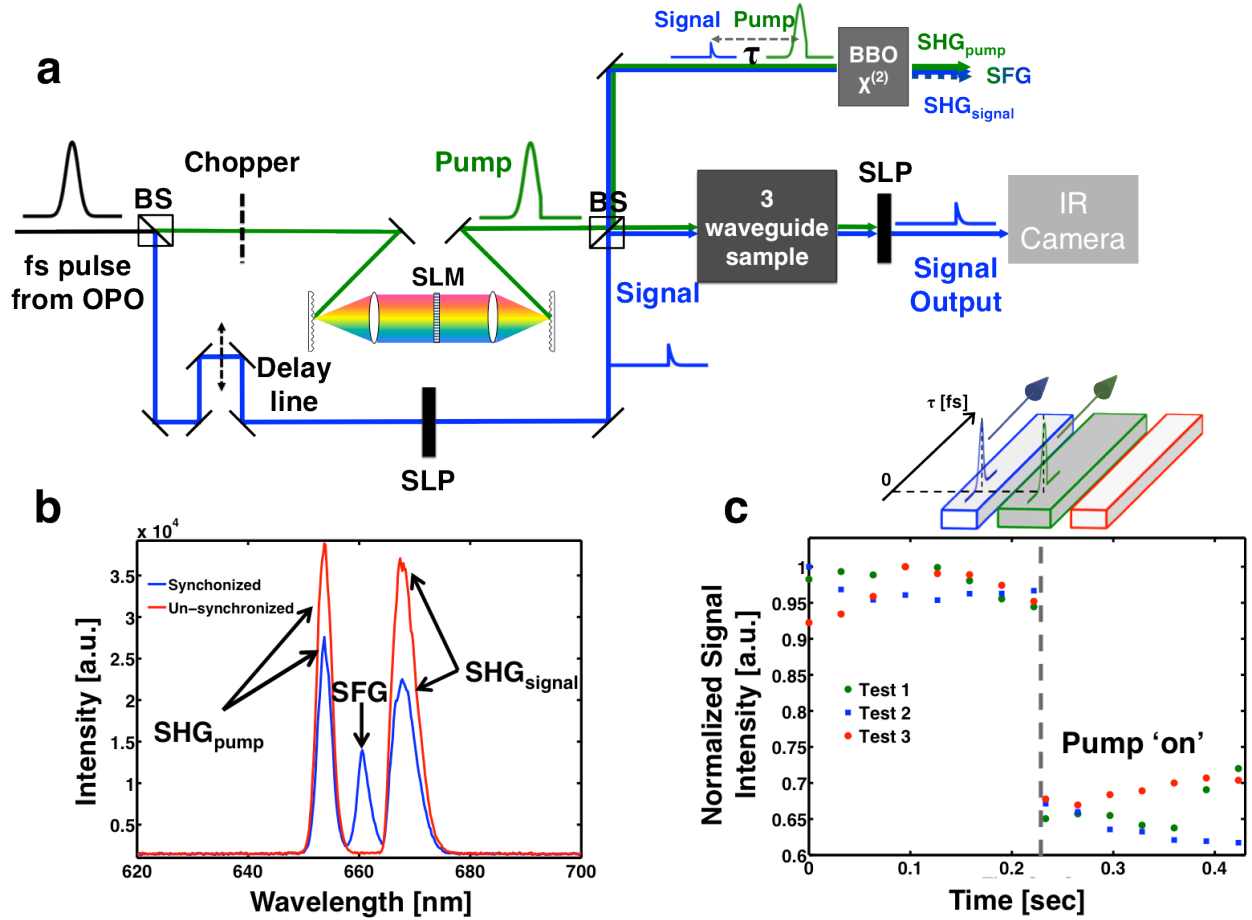

**Figure 4 – Active coupling control between outer waveguides in AE configuration. (a)** Schematic of the experimental apparatus. The ultrashort pulse (~140 fs with central wavelength at 1310nm) is split into a strong truncated pump beam that is coupled to the middle waveguide and the weak signal beam that is coupled to one of the outer waveguide. On the pump path, a mechanical chopper is used to turn off and on the injection of the pump pulse to the middle waveguide. On the signal path, a delay line is used to synchronize the entrance of the pump and signal. The signal is cut witg a sharp long pass (SLP) filter such it does not overlap with the pump. The synchronization between the pump and signal is achieved by focusing the beams onto a BBO crystal. **(b)** When the two beams are synchronized a Sum Frequency Generation (SFG) is observed (blue curve) in addition to each beam own Second Harmonic Generation (SHG) (red curve), a detailed procedure is found in the SOM (see S5). **(c)** Observed modulation as a result of varied coupling in the presence of a synchronized pump beam, showing a significant reduction of the signal intensity when the pump is present.

As we have demonstrated, this unique scheme separates spatially the control in the middle waveguide from the signal in the outer waveguides. Therefore, other physical effects can be employed to change the mode index of the middle waveguide such as thermal, mechanical or electronic processes. Finally, due to the fact that in AE evolution, the effective propagation length in the middle waveguide is reduced by several orders of magnitude, the AE scheme could also work when the middle waveguide is lossy, such as in the case of a plasmonic middle waveguide, where propagation constant and coupling coefficients are complex valued (See S1). This will allow to further reduce the footprint without facing the losses typically associated with plasmonic waveguides.

Inspired by the atomic AE, our scheme of active control for nanoscale waveguiding brings new insight for subwavelength Silicon photonics. Since AE enables unprecedented tunable coupling, it further allows a zero effective coupling between the outer waveguides that cancels light transfer and eliminates the cross talks between waveguides, a must for ultra-dense nano-photonics interconnects. This is due to a degeneracy in the coupled system, where the sequential coupling strength ($V_{12}V_{23}/\Delta\beta$) is equal to the direct coupling ($V_{13}$), leading to a full destructive interference along the entire propagation. This degeneracy happens when the middle waveguide's effective index is higher than the outer waveguides, i.e when $\Delta\beta = \Delta\beta_{23} = -\Delta\beta_{12} > 0$. We found that this constraint can be satisfied for TE mode where the sizes of the waveguides are of the same order as the gap between the waveguides, thus allowing significant coupling between the outer waveguides. Hence, at this singular point, all three waveguides are decoupled from each other despite the fact they are densely packed with a sub-wavelength gap (see S1), thus yielding the capability to shield information from the surrounding within the evanescent range. Therefore, AE scheme provides a new way in achieving dense optical waveguiding with negligible cross-talk. For example, for a given gap between two waveguides of 800 nm, coupling length under AE conditions can be as long as 1.3 mm, in sharp contrast with 105μm in a conventional directional coupler with the same gap. However, according to our simulations, dimensional accuracy required for the realization of such excellent shielding is on the order of a few nanometers, a challenging fabrication task at present. It is nonetheless possible as the nanofabrication further improves.

In conclusion we have experimentally demonstrated the AE decomposition scheme in nanowaveguides. We have shown that this approach enables on demand dynamical control of the coupling between two closely packed waveguides, by modulating the mode index of an in-between decoupled waveguide. This is in contrast to the conventional directional couplers where the constant coupling coefficient is pre-determined by the distance in between the waveguides. AE offers a unique route for optical information transfer in densely packed nano-scale photonic components, paving the way toward compact modulators, ultrafast optical signal routers and interconnects that will be the foundation of ultra-dense integrated nanophotonics.

# Methods

**Numerical simulations:** The design of the samples was carried out using semi-analytical calculations and COMSOL multiphysics simulations aimed to examine the generality of the AE scheme. We have also studied the dynamics in the modulation experiment by performing numerical simulation of the coupled nonlinear Schrodinger equations using finite difference calculation. In our simulations, we have used the nonlinear parameters of Silicon [23] ($n_{kerr} = 1.7 \cdot 10^{-5}\ cm^2/GW$) and estimated the pump intensity in the middle waveguide to be ($I_{pump} = 430\ GW/cm^2$).

**Sample preparation:** Silicon waveguides were fabricated in a silicon-on-insulator (SOI) substrate. The thickness of silicon and buried oxide were 340 nm and 1 um, respectively. A 160-nm-thick hydrogen silsesquioxane (HSQ) resist (Dow Corning XR-1541) was spun on the SOI substrate. The HSQ layer was patterned by electron-beam lithography for the etching mask. The silicon waveguides were formed by reactive ion etching (RIE) in Cl2/HBr/O2. 10:1 buffered HF was used to remove the HSQ mask.

The light propagation in the waveguides was examined both with far field and near field setups described below:

**Far Field measurements**: The far field measurements, presented in Fig.2, were carried out on microscope with a 10x/NA 0.3 objective in reflection mode. The light from a CW laser diode at 1310 nm is focused on the input grating. We collect the output light from the waveguides in reflection and image them an uncooled infrared focal plane array (IR FPA, Goodrich). The waveguides roughness scatters the coupled light along the propagation and serves as an indication of proper coupling [27].

**Near-Field Scanning optical microscopy (NSOM) measurements**: we acquired the NSOM images in Fig. 2 and in S3 using a home-built scattering near field optical microscopy (s-NSOM) [28] in a transmission configuration. In this setup, light from a CW laser diode at 1310 nm is focused on the input grating using a 10x/NA 0.3 objective. To ease the alignments, we used a PtIr coated Advanced Tip at the End of Cantilever (ATEC) with a tip radius < 10 nm and tapping frequency of Ω. The ATEC was moved along the waveguides with sub-nanometer precision using piezostages to record the propagation of the input light. Light scattered from and modulated by the tip was collected using an aspheric lens (NA=0.16) and focused onto a 10 MHz InGaAs photoreceiver (NewFocus 2053-FS). The signal was demodulated at 3Ω to filter out the non-near field background scattering.

**Ultrafast modulation measurements:** the ultrafast experimental setup, shown schematically in Fig. 4a, is based on a variant of the single-shot nonlinear technique [26] where an ultrashort pulse from a Ti:Sa pumped OPO (Coherent/APE Chameleon Compact OPO) centered at 1310 nm is split into strong pump beam that is coupled to the middle waveguide and weak signal beam that is coupled to one of the outer waveguide (for convenience we note it as waveguide #1). To ensure that the signal measured only arises from the light coupled to the outer waveguide, we spectrally truncate at 1319 nm the red tail of the pump beam using a pulse shaper while leaving spectrally untouched the signal beam. The truncated pump beam is coupled to the middle waveguide with a grating coupler. The signal beam is passed through a sharp long pass filter (Semrock 1319 nm RazorEdge® ultrasteep long-pass edge filter) before being injected waveguide#1 in a similar fashion. As the evolution in the AE configuration confines the pump beam to the middle waveguide along the propagation, it prevents leakage to the outer waveguides. The method used to ensure that the pump and signal beams are coupled in the waveguides at the exact same time (with an accuracy of 4 fs) is described in the supplementary material. The light propagated in the outer waveguide is out-coupled to the free space and imaged on an uncooled IR FPA (Goodrich), after passing through a sharp long pass filter (Semrock 1319 nm RazorEdge® ultrasteep long-pass edge filter). The wavelengths contributing to the image formed on the camera are spectrally located in the tail that has been cut away from the pump but left in the signal. With a mechanical chopper we turn on and off the pump beam while recording second the images from the IR FPA at 31 frames per second. We processed the acquired images to extract the output from the signal waveguide as a function of the pump.